\documentclass[useAMS,usenatbib,a4paper]{mn2e}
\usepackage[a4paper,colorlinks=True,citecolor=blue]{hyperref}
\usepackage{amssymb}
\usepackage{graphicx}
\usepackage{epstopdf}
\newcommand{\bea}{\begin{eqnarray}}
\newcommand{\eea}{\end{eqnarray}}
\newcommand{\beq}{\begin{equation}}
\newcommand{\eeq}{\end{equation}}

\newcommand{\zobov}{{\small ZOBOV} }

\title[Tracing $\Phi$ with cosmic voids]{Tracing the gravitational potential using cosmic voids}
\author[S. Nadathur et al.]{Seshadri Nadathur$^{1}$\thanks{seshadri.nadathur@port.ac.uk}, Shaun Hotchkiss$^{2,3}$ and Robert Crittenden$^{1}$\\
$^1$Institute of Cosmology and Gravitation, University of Portsmouth, Burnaby Road, Portsmouth PO1 3FX, UK\\
$^2$Department of Physics, University of Auckland, Private Bag 92019, Auckland, New Zealand\\
$^3$Department of Physics and Astronomy, University of Sussex, Falmer, Brighton, BN1 9QH, UK
}

\begin{document}

\date{\today}

\pagerange{\pageref{firstpage}--\pageref{lastpage}}

\label{firstpage}

\maketitle

\begin{abstract}
The properties of large underdensities in the distribution of galaxies in the Universe, known as cosmic voids, are potentially sensitive probes of fundamental physics. We use data from the MultiDark suite of $N$-body simulations and multiple halo occupation distribution mocks to study the relationship between galaxy voids, identified using a watershed void-finding algorithm, and the gravitational potential $\Phi$. We find that the majority of galaxy voids correspond to local density minima in larger-scale overdensities, and thus lie in potential wells. However, a subset of voids can be identified that closely trace maxima of the gravitational potential and thus stationary points of the velocity field. We identify a new void observable, $\lambda_v$, which depends on a combination of the void size and the average galaxy density contrast within the void, and show that it provides a good proxy indicator of the potential at the void location. A simple linear scaling of $\Phi$ as a function of $\lambda_v$ is found to hold, independent of the redshift and properties of the galaxies used as tracers of voids. We provide an accurate fitting formula to describe the spherically averaged potential profile $\Phi(r)$ about void centre locations. We discuss the importance of these results for the understanding of the evolution history of voids, and for their use in precision measurements of the integrated Sachs--Wolfe effect, gravitational lensing and peculiar velocity distortions in redshift space. 
\end{abstract}


\section{Introduction}
\label{sec:introduction}

Large underdensities in the matter distribution of the Universe, known as cosmic voids, are interesting objects that can be used as probes of cosmology and fundamental physics. This is because their properties are sensitive to both the initial conditions of the density perturbations and to the growth rate of structure.

Several catalogues of cosmic voids have been compiled from galaxy redshift survey data \citep{Pan:2012,Sutter:2012wh,Nadathur:2014a,Nadathur:2016a,Mao:2016a}, which can be used in observational tests. The simplest of these tests concern the number of voids and their size distributions, which may already provide constraints on the dark energy equation of state and modified gravity scenarios \citep[e.g.][]{Li:2012,Clampitt:2013,Cai:2015,Pisani:2015,Nadathur:2016a}. 

Voids induce small secondary anisotropies in the cosmic microwave background (CMB) via the integrated Sachs-Wolfe (ISW) effect. Numerous studies have attempted to measure this effect \citep[e.g.][]{Granett:2008a,Ilic:2013,Cai:2014,Hotchkiss:2015a,Granett:2015,Nadathur:2016b,Cai:2016b,Kovacs:2016}, using different methods of varying statistical power, and reaching widely varying conclusions. Voids also produce measurable weak lensing signals \citep[e.g.][]{Krause:2013}. Void lensing of background source galaxies has been studied by \citet{Melchior:2014,Clampitt:2015,Sanchez:2016}, and recently \citep{Cai:2016b} made a measurement of CMB lensing by voids. The anisotropic distortion of void shapes in redshift space \citep{Ryden:1995} can be used to test the expansion history of the Universe \citep[e.g.][]{Lavaux:2012,Mao:2016b}, via a form of the Alcock-Paczynski (AP) test \citep{Alcock:1979}. 

Both the lensing and ISW effects of voids are dependent on the larger-scale density environment they are located in, which is reflected in the gravitational potential $\Phi$. The AP distortion test is also indirectly sensitive to this environment due to the complicating effects of peculiar velocity flows around the void. Beyond its implications for the use of voids as observational tools, the gravitational potential environment in which voids are located is of intrinsic interest due to its role in the formation and evolution of the Cosmic Web.

A complication to understanding void environments is the fact that in practice voids are observed simply as underdense regions in the distribution of bright galaxies, whereas the properties of the environment depend on the distribution of dark matter. Unfortunately, existing analytic models of void formation and growth \citep{Sheth:2003py} and their extension to galaxy voids \citep{Furlanetto:2006} fail to correctly describe the properties of practically observable voids \citep[for a discussion, see][]{Nadathur:2015b}. An understanding of the relationship between voids and the potential $\Phi$ must therefore be built on the results from $N$-body simulations populated with realistic mock galaxies: insights from such studies may then be used to develop new theoretical models of voids.

Our aim in this paper is to provide such a study. We aim to answer the following questions. Can voids in the galaxy distribution be used to trace features (in particular, points of maxima) in the gravitational potential, $\Phi$? Can the average value of $\Phi$ at void locations, and its local variation around these points, be predicted on the basis of easily observable properties of galaxy voids? How does this relationship depend on the properties (e.g., the luminosity) of the galaxy population used to identify voids and the redshift of observation, and can any universal trends be identified? Can such knowledge of the properties of $\Phi$ be used to improve the sensitivity of methods to measure the ISW and lensing effects of voids, and to understand the impact of peculiar velocity flows on voids observed in redshift space? 

To answer these questions we make use of different mock galaxy populations at different redshifts within a large $N$-body simulation in which the true dark matter density and $\Phi$ are both known. We identify voids using the \zobov watershed algorithm \citep{Neyrinck:2008}, which can be robustly adapted for use with observational data and has been used in the construction of most of the existing void catalogues discussed above. We show that while the majority of voids identified using this algorithm are merely local density minima in otherwise contracting regions, an identifiable subset of voids does closely trace points of maxima of $\Phi$. We identify a combination of void properties encapsulated in a new void parameter $\lambda_v$, which is an excellent and universal indicator of the value of $\Phi$, and provide fitting formulae for the mean spherically averaged profiles $\Phi(r)$ about void locations. We relate the gravitational potential environment to void density profiles and the compensation of the mass deficit within voids, and discuss their redshift evolution. Finally, we also discuss an extension of our results to the case of `superclusters': large-scale overdense structures analogous to the void underdensities.

The layout of the paper is as follows. In Section \ref{sec:numerical} we describe the properties of the simulation, construction of the mock galaxy populations and the identification of voids. In Section \ref{sec:voidPhi} we investigate the relationship between these voids and the gravitational potential and lay out our results. In Section \ref{sec:applications} we discuss some applications of these results to particular observational studies using voids. We conclude in Section \ref{sec:discussion}. Numerical details for the various fitting formulae provided are summarised in the appendices, where we also provide some additional comparisons of different methods of locating voids.

\section{Numerical methods}
\label{sec:numerical}

\subsection{Simulations}
\label{subsec:sims}

We made use of the Big MultiDark (BigMD) $N$-body simulation \citep{Klypin:2016} from the MultiDark simulation project \citep{Prada:2012}. This simulation follows the evolution of $3840^3$ particles in a box of side $L=2.5\;h^{-1}$Gpc using the {\small GADGET-2} \citep{Springel:2005} and Adaptive Refinement Tree (ART) \citep{Kravtsov:1997,Gottloeber:2008} codes, with cosmological parameters $\Omega_M=0.307$, $\Omega_B=0.048$, $\Omega_\Lambda=0.693$, $n_\rmn{s}=0.95$, $\sigma_8=0.825$ and $h=69.3$. Initial conditions for the simulation were set using the Zeldovich approximation at starting redshift $z_i=100$.

We used simulation data from three different redshift snapshots, $z=0.1$, $z=0.32$ and $z=0.52$. These snapshots were chosen to be as close as possible to the median redshifts of the galaxy populations we wish to model, as described below. On each snapshot, we used catalogues of haloes found using the Bound Density Maximum algorithm \citep{Klypin:1997,Riebe:2013}. To measure densities, we used the underlying DM density field determined from the full resolution simulation output on a $2350^3$ grid using a cloud-in-cell interpolation at each of these three redshifts. This density field was then smoothed using a Gaussian kernel of width equal to one grid cell.

The gravitational potential $\Phi$ is related to the DM density field $\delta(\mathbf{r})$ by the Poisson equation 
\beq
\label{eq:Poisson}
\nabla^2 \Phi(\mathbf{r},z) = \frac{3}{2}\frac{H_0^2\Omega_M}{a(z)}\delta(\mathbf{r},z)\,.
\eeq
In Fourier space this can be written conveniently as 
\beq
\label{eq:PoissonF}
\Phi(\mathbf{k},z) = \frac{3}{2}\frac{H_0^2\Omega_M}{a(z)k^2}\delta(\mathbf{k},z)\,.
\eeq
We solved this equation numerically on the grid at each redshift to obtain $\Phi$. 

Due to the $k^{-2}$ factor in equation \ref{eq:PoissonF}, $\Phi$ varies on much larger scales than $\delta$. We therefore downgraded the $\Phi$ obtained to a $1175^3$ grid (i.e., a resolution of $\sim2\;h^{-1}\mathrm{Mpc}$) and applied a Gaussian smoothing of one grid cell width to ease subsequent computational requirements. This means that we are not sensitive to peaks of $\Phi$ occurring on scales smaller than our smoothing scale. For the void sizes we obtain below, such small peaks are not observationally relevant.

\begin{table*}
\begin{minipage}{130mm}
\caption{Properties of the mock HOD samples used in this work. }
\begin{tabular}{@{}ccccccccc}
\hline
Sample name & Redshift &$\log M_\rmn{min}$ & $\sigma_{\log M}$
& $\log M_0$ & $\log M_1^\prime$ & $\alpha$ & $\overline{\rho_g}$ & $b_g$ \\
& & & & & & & $ (h^3\rmn{Mpc}^{-3})$ &\\
\hline
Main1 & 0.10 & 12.14 & 0.17 & 11.62 & 13.43 & 1.15 & $3.18\times 10^{-3}$ & 1.3 \\
Main2 & 0.10 & 12.78 & 0.68 & 12.71 & 13.76 & 1.15 & $1.16\times 10^{-3}$ & 1.4 \\
LOWZ & 0.32 & 13.20 & 0.62 & 13.24 & 14.32 & 0.93 & $2.98\times 10^{-4}$ & $\sim2$ \\
CMASS & 0.52 & 13.09 & 0.596 & 13.08 & 14.00 & 1.013 & $2.0\times 10^{-4}$ & $\sim2$ \\
\hline\\
\end{tabular}
\label{table:HODsamples}
\end{minipage}
\end{table*}

\subsection{HOD modelling of galaxy tracers}
\label{subsec:HOD}

To create mock galaxy catalogues to use as tracers of voids, we populated simulation haloes according to the Halo Occupation Distribution (HOD) model of \citet{Zheng:2007}, assigning galaxies to a DM halo according to a distribution based on the halo mass $M$. Central and satellite galaxies were treated separately. The mean occupation function of central galaxies was parametrised as
\beq
\label{eq:Ncen}
\left<N_\rmn{cen}(M)\right>=\frac{1}{2}\left[1+\rmn{erf}\left(\frac{\log M-\log M_\rmn{min}}{\sigma_{\log M}}\right)\right]\,,
\eeq
and the number of central galaxies in each mass bin follows a nearest-integer distribution. The number of satellite galaxies follows a Poisson distribution with
\beq
\label{eq:Nsat}
\left<N_\rmn{sat}(M)\right>=\left<N_\rmn{cen}(M)\right>\left(\frac{M-M_0}{M_1^\prime}\right)^\alpha\,.
\eeq
Central galaxies were placed at the centre of their respective haloes, while satellite galaxies were distributed through the halo with radial distances from the centre drawn from a random distribution based on a fiducial Navarro-Frenk-White (NFW) mass profile \citep{NFW:1996,NFW:1997}.

The five parameters $M_\rmn{min}$, $M_0$, $M_1^\prime$, $\sigma_{\log M}$ and $\alpha$ of this HOD model were chosen in order to match those determined in the literature for different galaxy populations. Two of our mock samples, labelled Main1 and Main2, were obtained from the $z=0.1$ snapshot data and were designed to match two low-redshift volume-limited luminosity threshold samples from the Sloan Digital Sky Survey (SDSS) Data Release 7 (DR7) Main galaxy samples \citep{Zehavi:2011}. These two samples allow us to isolate the effects of mean galaxy density and bias from the redshift evolution. We used the $z=0.32$ snapshot data to model a population matching the properties of the SDSS Baryon Oscillation Spectroscopic Survey (BOSS) LOWZ galaxies \citep{Manera:2015} and the $z=0.52$ data to model the BOSS CMASS galaxies \citep{Manera:2013}. The parameters used in each case, as well as the average number density, $\overline{\rho_g}$, and approximate bias, $b_g$, of the original galaxy samples, are summarised in Table~\ref{table:HODsamples}.

%==================Fig.: =======================%
\begin{figure}
\begin{center}
\includegraphics[width=85mm]{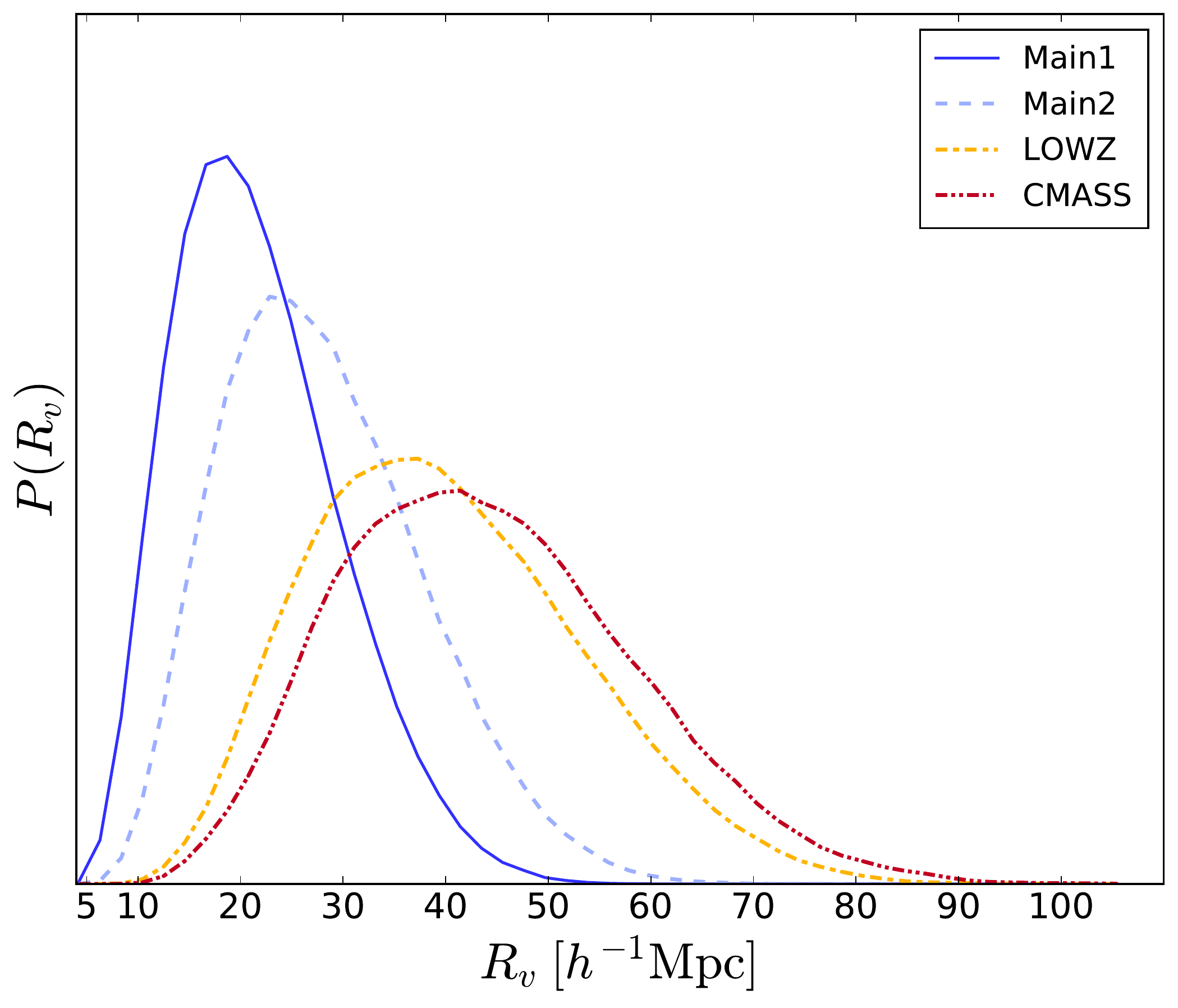}
\caption{Normalized comoving size distributions for voids in each of the four galaxy populations considered in this work. Sparser and more highly biased galaxy tracers result in larger voids.} 
\label{fig:voidprops}
\end{center}
\end{figure}
%==================Fig.:=======================%

\subsection{Void finding}
\label{subsec:void-finding}

To identify voids in the galaxy mocks described above, we used the \zobov watershed void-finding algorithm \citep{Neyrinck:2008}. \zobov uses a Voronoi tessellation field estimator (VTFE) technique to reconstruct the tracer density field from a discrete distribution of particles. It then identifies local minima in this field and the watershed basins around them, which form a non-overlapping set of density depressions or voids. In identifying these voids, the algorithm makes no prior assumptions about the void shape, instead respecting the true topology of underdensities in the galaxy distribution. To a very good approximation our void-finding algorithm is \emph{space-filling}, i.e. the total volume of all voids equals that of the simulation box. This is a common feature of most watershed-based void finding algorithms.

After the identification of all local density minima, it is possible to merge some neighbouring minima together to form larger voids, and variations of this step have often been performed in the literature. However, any such merging procedure is always based on subjective criteria \citep{Nadathur:2015c,Nadathur:2016a}. It also reduces the number of unique voids available for statistical analyses, and obscures some useful degeneracies between void properties. We therefore followed the procedure of \citet{Nadathur:2016a} by defining each individual density basin as a distinct void, without any additional merging. 

We used the following key observable properties of each void, which can be determined from the galaxy distribution alone:
\begin{enumerate}
\item the location of the void centre, $\mathbf{X}_v$, defined to be the centre of the largest completely empty sphere that can be inscribed within the void \citep{Nadathur:2015b,Nadathur:2016a}, 
\item the effective void size $R_v= \left(3V/4\pi\right)^{1/3}$, where the total void volume $V$ is determined from the sum of the volumes of its constituent Voronoi cells,
\item the minimum galaxy density within a void, $\delta_{g,\rmn{min}}=\rho_{g,\rmn{min}}/\overline{\rho_g}-1$, where $\rho_{g,\rmn{min}}$ is the minimum VTFE reconstructed galaxy density in the void, 
\item the \emph{average} galaxy density contrast over the void, $\overline\delta_g \equiv \frac{1}{V}\int_V\frac{\rho_g(\mathbf{x})}{\overline{\rho_g}}\,\rmn{d}^3\mathbf{x} -1$, which is in practice estimated from the volume-weighted average density of the void Voronoi cells, $\overline\delta_g = \frac{1}{\overline{\rho_g}}\frac{\sum_i \rho_g^iV_i}{\sum_iV_i}-1$,
\item and the \emph{density ratio} $r=\rho_{g,\rmn{ridge}}/\rho_{g,\rmn{min}}$, defined as the ratio of the lowest value of the galaxy density along the edge of the void's watershed basin to the minimum galaxy density at the void centre \citep{Neyrinck:2008}.
\end{enumerate}
Note that with the exception of $\mathbf{X}_v$ and $\delta_{g,\rmn{min}}$, all other void properties would change if merging of neighbouring voids were allowed.

In Appendix \ref{appendix:centres} we examine the effect of a common alternative definition of the void centre location based on the volume-weighted barycentre of void galaxies, and show that it is an inferior tracer of the gravitational potential. We also briefly discuss the effects of merging neighbouring voids together using different prescriptions.

%==================Fig.: =======================%
\begin{figure*}
\begin{center}
\includegraphics[width=160mm]{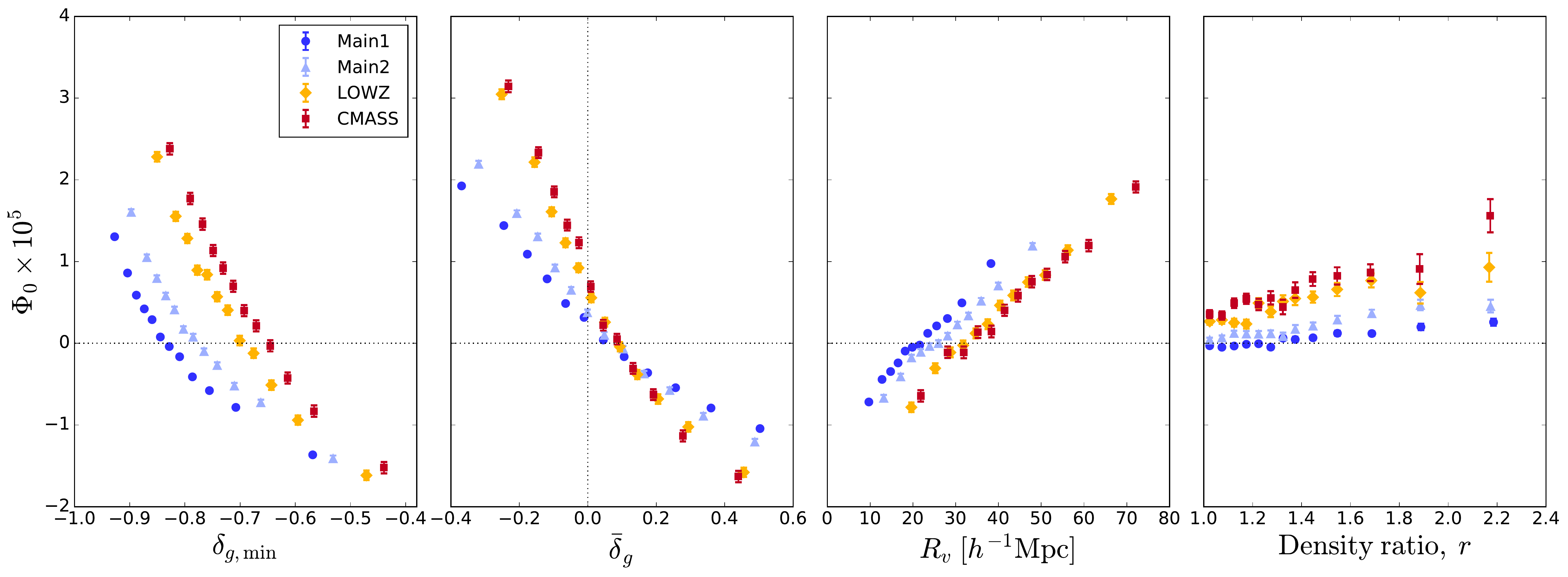}
\caption{Variation of the average gravitational potential at the void centre, $\Phi_0$, with changes in: the minimum galaxy density contrast within the void, $\delta_{g,\rmn{min}}$; the average galaxy density contrast over the void volume, $\overline\delta_g$; void size, $R_v$; and void density ratio $r$. Data points show the bin mean values, and bars represent the standard error in these means. The strongest correlation is with $\overline\delta_g$, with $\overline\delta_g\simeq0.1$ also representing a universal turnover point between $\Phi_0>0$ and $\Phi_0<0$ for voids in all galaxy tracers.}
\label{fig:Phiobservables}
\end{center}
\end{figure*}
%==================Fig.:=======================%

\section{Voids and the gravitational potential}
\label{sec:voidPhi}

Differences in the bias and mean number density of tracer galaxy populations lead to differences in the characteristics of the voids obtained from them \citep{Sutter:2014b,Nadathur:2015c}. This is true even when the different galaxy types trace exactly the same underlying density field, as is the case for our Main1 and Main2 samples at $z=0.1$. In sparser galaxy samples the resolution of the VTFE density reconstruction is reduced, so smaller voids are either not seen at all, or subsumed within larger neighbours. The lower resolution also inhibits the precision with which the locations of density minima can be identified, so voids in sparser galaxy samples are on average shallower. However, within the population of voids from any one galaxy sample, the principle of the watershed void-finding algorithm means that larger voids correspond to deeper density minima \citep{Nadathur:2015b,Nadathur:2015c}. Figure~\ref{fig:voidprops} shows the resulting size distribution for voids in each of the four galaxy samples. 

\subsection{Inferring $\Phi$ from void observables}
\label{subsec:inferringPhi}
 
Almost all of the voids we identified correspond to genuine underdensities in the DM distribution in the simulation box \citep[the false positive identification rate for voids is no higher than 3\%; see][]{Nadathur:2016a}. Naively, one might infer an association between underdensities in the matter distribution and regions with $\Phi>0$. However, even at late times, to a good approximation only half the volume of the Universe corresponds to regions of $\Phi>0$, in contrast to the case for the dark matter density field. Given that watershed voids are space-filling, this naive association clearly cannot hold for all voids.  

We therefore measured the value of the gravitational potential at the void centre, $\Phi_0 = \Phi(\mathbf{X}_v)$, for each void in the different populations. As expected, almost $50\%$ of the voids in the Main1 sample correspond to values of $\Phi_0<0$. This fraction drops slightly with decreasing sparsity, to $\sim45\%$ of voids in the CMASS sample.

This distinction between $\Phi_0>0$ and $\Phi_0<0$ points to a fundamental difference in void environments. Voids in regions of $\Phi<0$ are local density minima lying within large-scale overdensities. Such voids will eventually be crushed out of existence by the infall of matter from their surroundings, culminating in the \emph{void-in-cloud} scenario described by \citet{Sheth:2003py}. On the other hand, those corresponding to $\Phi>0$ will continue to expand outwards becoming ever emptier. In practical terms, the measurement of gravitational effects such as the integrated Sachs-Wolfe (ISW) anisotropies due to voids or the redshift-space distortions (RSD) associated with their expansion require a statistically robust distinction to be made between the two sub-populations on the basis of their observable properties.

Figure~\ref{fig:Phiobservables} shows trends in the binned average values of $\Phi_0$ as functions of the four void observables introduced in Section~\ref{subsec:void-finding}. Similar trends are seen for each galaxy tracer type, but voids in more highly biased tracers and in samples at higher redshift are on average associated with larger values of $\Phi_0$. The slopes of the trend lines in each panel indicate the discriminatory power of each observable as a proxy for the void environment. The average galaxy density $\overline\delta_g$ is the single best predictor of the value of $\Phi_0$. Even more interestingly, the point of zero-crossing is the same for voids in each of the four mock galaxy samples, occurring at $\overline{\delta}_g\simeq0.1$ independent of the sparsity or galaxy bias. The value of $\overline{\delta}_g$ thus provides a \emph{universal} indicator of the large-scale void environment, which can be used for voids traced by all galaxy types. This universality is closely related to the results of \citet{Nadathur:2015c}, who showed that $\overline{\delta}_g$ is also a universal indicator of the large-scale density compensation around void locations.

There is also a clear trend for $\Phi_0$ to increase with void depth (lower values of $\delta_{g,\rmn{min}}$), as intuitively expected. However, the correlation is smaller than for $\overline\delta_g$, and the threshold value of $\delta_\rmn{g,min}$ separating $\Phi_0>0$ from $\Phi_0<0$ depends on the bias, sparsity and redshift of the tracer galaxy sample, limiting the usefulness of this observable. A somewhat weaker correlation is also seen with the void size $R_v$, which follows from the fact that $R_v$ is itself strongly correlated with $\delta_{g,\rmn{min}}$ \citep{Nadathur:2015b,Nadathur:2015c}.

Significantly, we found that the density ratio $r$ shows little correlation with $\Phi_0$. This follows the finding of \citet{Nadathur:2016a} that the density ratio also shows very little correlation with the void significance. Although some previous studies have used the density ratio to define selection criteria for void populations used in ISW detection \citep[e.g.][]{Granett:2008a,Cai:2016b} or void RSD effects \citep{Mao:2016b}, our results show that there is no theoretical justification for such a strategy.

%==================Fig.: =======================%
\begin{figure}
\begin{center}
\includegraphics[width=85mm]{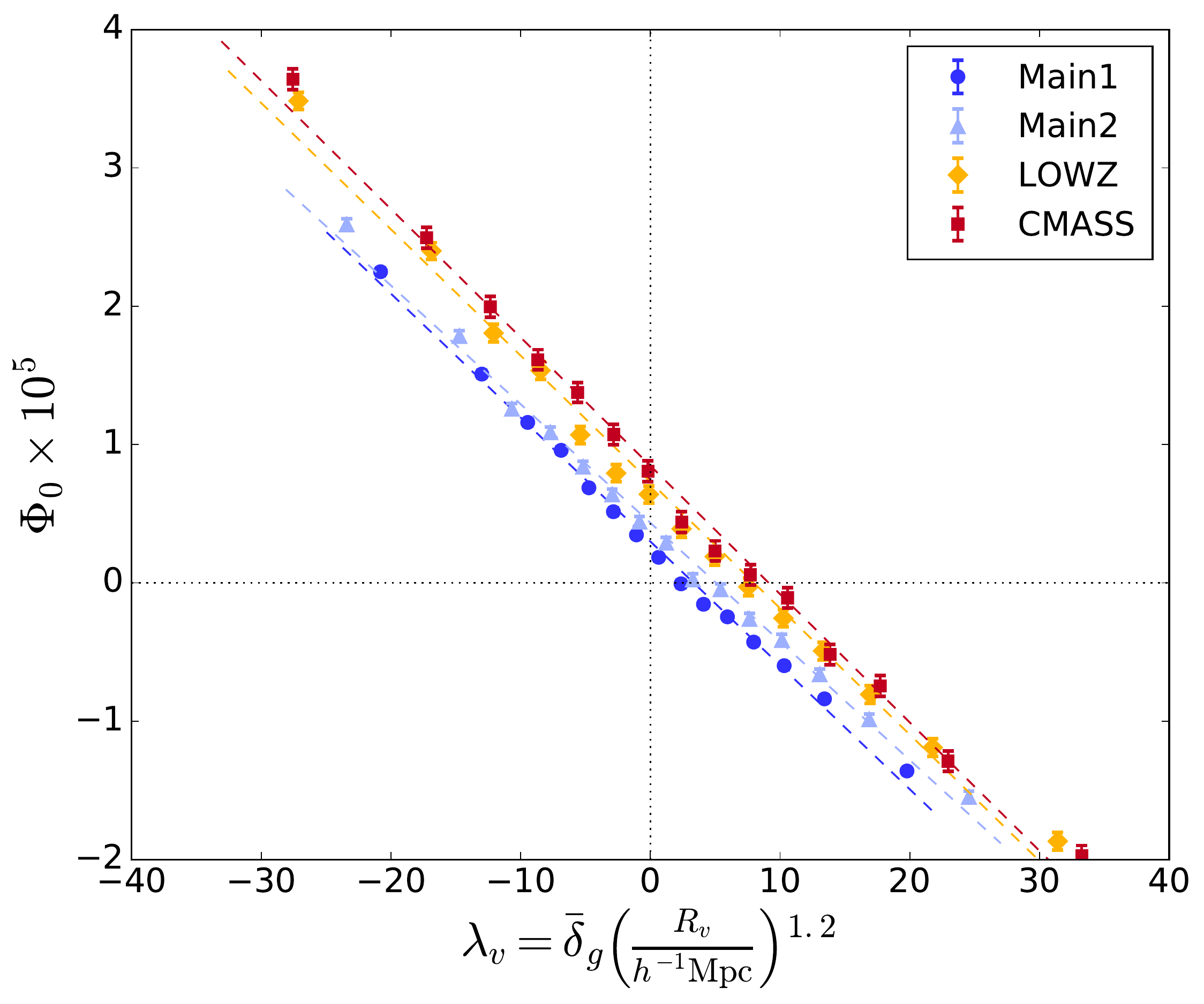}
\caption{Variation of the average gravitational potential at the void centre, $\Phi_0$, as a function of the void parameter $\lambda_v$, for voids in each of the four mock galaxy samples. A simple linear scaling is seen, and the slope of the best-fit line is very similar in each case, independent of tracer galaxy density, bias or redshift.} 
\label{fig:Phi_scaling}
\end{center}
\end{figure}
%==================Fig.:=======================%

\subsection{Scaling relations for $\Phi$}
\label{subsec:scaling}

Athough $\overline\delta_g$ is the single observable best correlated with the value of the gravitational potential at the void location, after accounting for this correlation there remains a residual dependence of $\Phi_0$ on other void parameters $\delta_{g,\rmn{min}}$ and $R_v$. Given that $\delta_{g,\rmn{min}}$ and $R_v$ are themselves strongly correlated with each other, it is enough to consider just one of them, which for convenience we took to be $R_v$. We found that for voids in all galaxy tracer samples, the average value of the potential at the void centre is extremely well fitted by the empirical formula
\beq
\label{eq:Phiscaling}
\overline\Phi_0(\overline\delta_g,R_v) = -a\lambda_v + c\,,
\eeq
where $a$ and $c$ are positive constants determined from the simulation and the parameter $\lambda_v$ is defined as
\beq
\label{eq:lambda_v}
\lambda_v = \overline\delta_g\left(\frac{R_v}{h^{-1}\rmn{Mpc}}\right)^{1.2}\,,
\eeq
determined from the measured values of $\overline\delta_g$ and $R_v$. Figure~\ref{fig:Phi_scaling} shows the best-fit forms of equation~\ref{eq:Phiscaling} for voids in each mock galaxy sample, and the fitted values of $a$ and $c$ are summarised in Appendix~\ref{appendix:fits}. The remarkable linear scaling of $\overline\Phi_0$ with $\lambda_v$ is universal for all tracer galaxy densities, bias values and redshifts. In addition, the slope of the scaling relation is also relatively independent of the galaxy tracer properties. For any individual void, the value of the potential gravitational at the void centre is a stochastic quantity drawn from a normal distribution with mean $\overline\Phi_0$. The variance of this distribution varied for the different galaxy populations, but was not found to have any significant dependence on $\lambda_v$. Numerical values are summarised in Table~\ref{table:Phi0fit}.

\subsection{Voids as tracers of maxima of $\Phi$}
\label{subsec:maxima}

%==================Fig.: =======================%
\begin{figure*}
\begin{center}
\includegraphics[width=150mm]{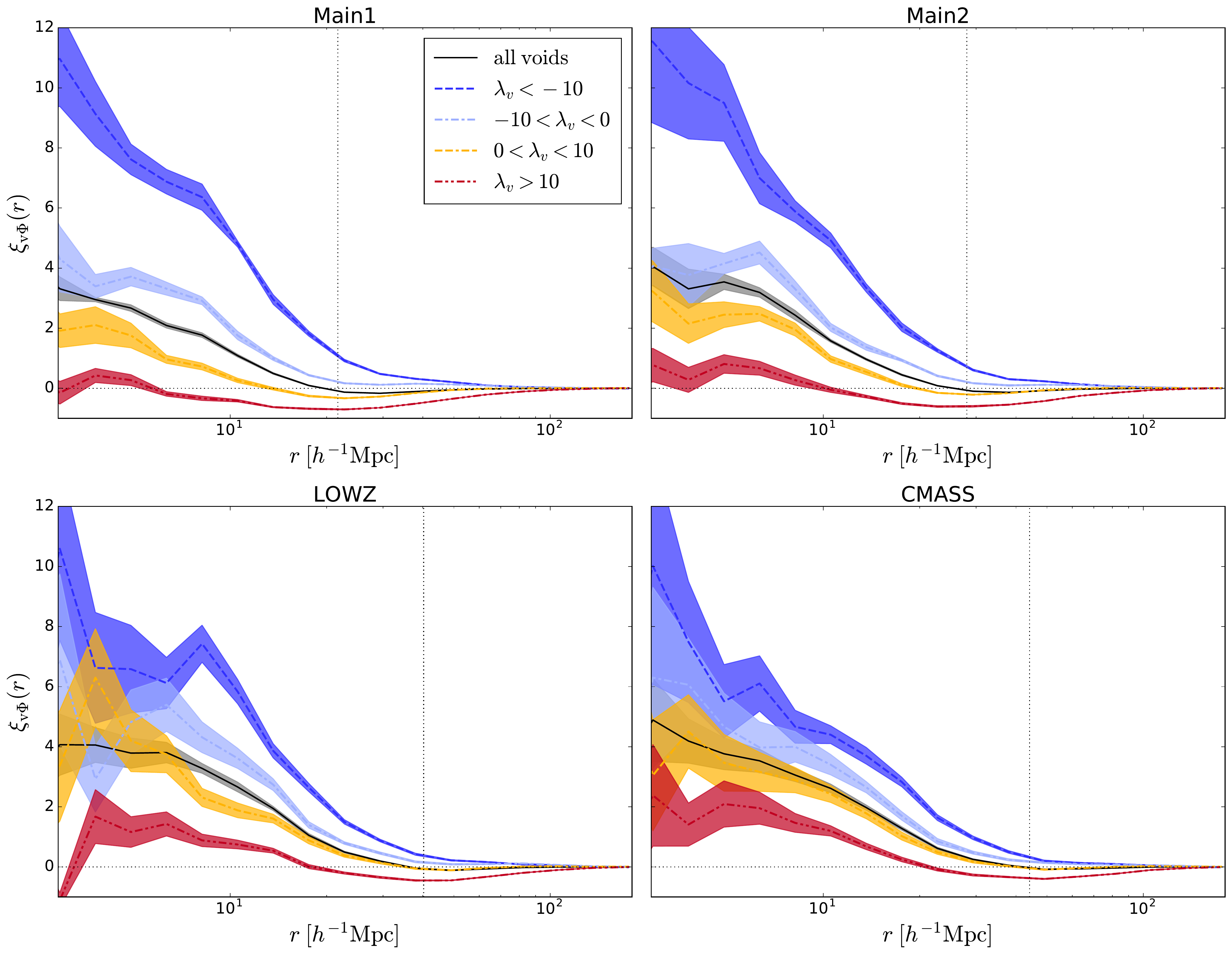}
\caption{The cross-correlation between positions of void centres and positions of maxima of the gravitational potential $\Phi$ as a function of distance, for voids in each mock galaxy sample. Shaded regions show the $68\%$ C.L. regions determined from jack-knife errors. In each panel, the solid (black) line shows $\xi_{v\Phi}(r)$ for all voids: this increases at small values of separation $r$, indicating that void locations do on average trace maxima of $\Phi$. Vertical dashed lines indicate the mean void size in each sample. The other lines in each panel show the same measurement for subsets of voids separated on the basis of the quantity $\lambda_v$ as indicated. The correlation increases markedly as the value of $\lambda_v$ decreases (becomes more negative).}
% Voids with $\lambda_v>10$, accounting for $\sim30\%$ of the total in each sample, show little or no correlation with maxima of $\Phi$ at small separations and are significantly anti-correlated at larger values of $r$.}} 
\label{fig:CCF_all}
\end{center}
\end{figure*}
%==================Fig.:=======================%

We next investigated the question of whether the locations of void centres correspond to locations of local maxima of the gravitational potential $\Phi$. To do this we first found all local maxima of $\Phi$ in each simulation slice, using a version of the \zobov watershed algorithm modified to work with the gridded $\Phi$ data. The location of each local maximum, $\mathbf{X}_\Phi$, could thus be determined to within a resolution of $\sim2h^{-1}$Mpc. Note that we recorded the position of all local maxima of the potential, including those for which $\Phi(\mathbf{X}_\Phi)<0$.

We then estimated the cross-correlation function (CCF) between the discrete sets of locations $\{\mathbf{X}_v\}$ and $\{\mathbf{X}_\Phi\}$ as a function of their separation $r$, using the Landy-Szalay estimator \citep{Landy:1993}
\beq
\label{eq:CCF}
\xi_{v\Phi}(r) = \frac{D_1D_2-D_1R-D_2R+RR}{RR}\,,
\eeq
where $D_1D_2$ is the (normalised) number of void centre-$\Phi$ maximum pairs separated by $r\pm \rmn{d}r/2$, $RR$ is the number of such pairs for a random point distribution, $D_1R$ is the number of void-random pairs and so on. To estimate the error on our measurement of $\xi_{v\Phi}$ we divided the simulation box up into $N=64$ sub-boxes and used a jack-knife procedure to obtain the variance
\beq
\label{eq:JKvar}
\sigma^2_\rmn{JK}\left(\xi_{v\Phi}^i\right) = \frac{(N-1)}{N}\times\sum_{\rmn{JK}-k=1}^N\left[\left(\xi_{v\Phi}^i\right)^{\rmn{JK}-k} -\overline{\xi_{v\Phi}^i}\right]^2\,,
\eeq
where $\left(\xi_{v\Phi}^i\right)^{\rmn{JK}-k}$ is the measured value of the $k$-th jack-knife realisation of $\xi_{v\Phi}$ in the $i$-th spatial bin, and 
\beq
\label{eq:JKmean}
\overline{\xi_{v\Phi}^i} = \frac{1}{N}\sum_{\rmn{JK}-k=1}^N\left(\xi_{v\Phi}^i\right)^{\rmn{JK}-k}
\eeq
is the mean over the jackknife samples.

The resulting estimates of the void-$\Phi$ CCF are shown in Figure~\ref{fig:CCF_all}. For voids in each galaxy sample $\xi_{v\Phi}$ is seen to peak at small separations $r$, showing that voids are indeed typically located near local maxima of $\Phi$. The identification is not perfect, and so the CCF remains positive out larger separations of $\mathcal{O}(10)\;h^{-1}$Mpc before turning over to $\xi_{\rmn{v}\Phi}<0$. 

Based on the results in the previous section, Figure~\ref{fig:CCF_all} also shows the result of the CCF measurement after splitting the void samples into bins based on the values of $\lambda_v$. In each case, a steady trend is seen towards increasing $\xi_{\rmn{v}\Phi}$ with decreasing $\lambda_v$, i.e. that voids with smaller (more negative) values of $\lambda_v$ are much more strongly correlated the positions of maxima of the gravitational potential. Voids with larger (positive) values of $\lambda_v$ show little or no evidence for $\xi_{\rmn{v}\Phi}(r)>0$ at small $r$, indicating that they are not associated with maxima of $\Phi$, and in fact show an anti-correlation at intermediate separation scales. This is particularly true for the $\lambda_v>10$ bin, which contains approximately $30\%$ of all identified voids in each sample. This trend with $\lambda_v$ is primarily driven by a similar trend with $\overline\delta_g$ --- the lower the average galaxy density within the void, the higher the likelihood that it is associated with a peak in the gravitational potential.

\subsection{$\Phi(r)$ profiles about void locations}
\label{subsec:profiles}

%==================Fig.: =======================%
\begin{figure*}
\begin{center}
\includegraphics[width=150mm]{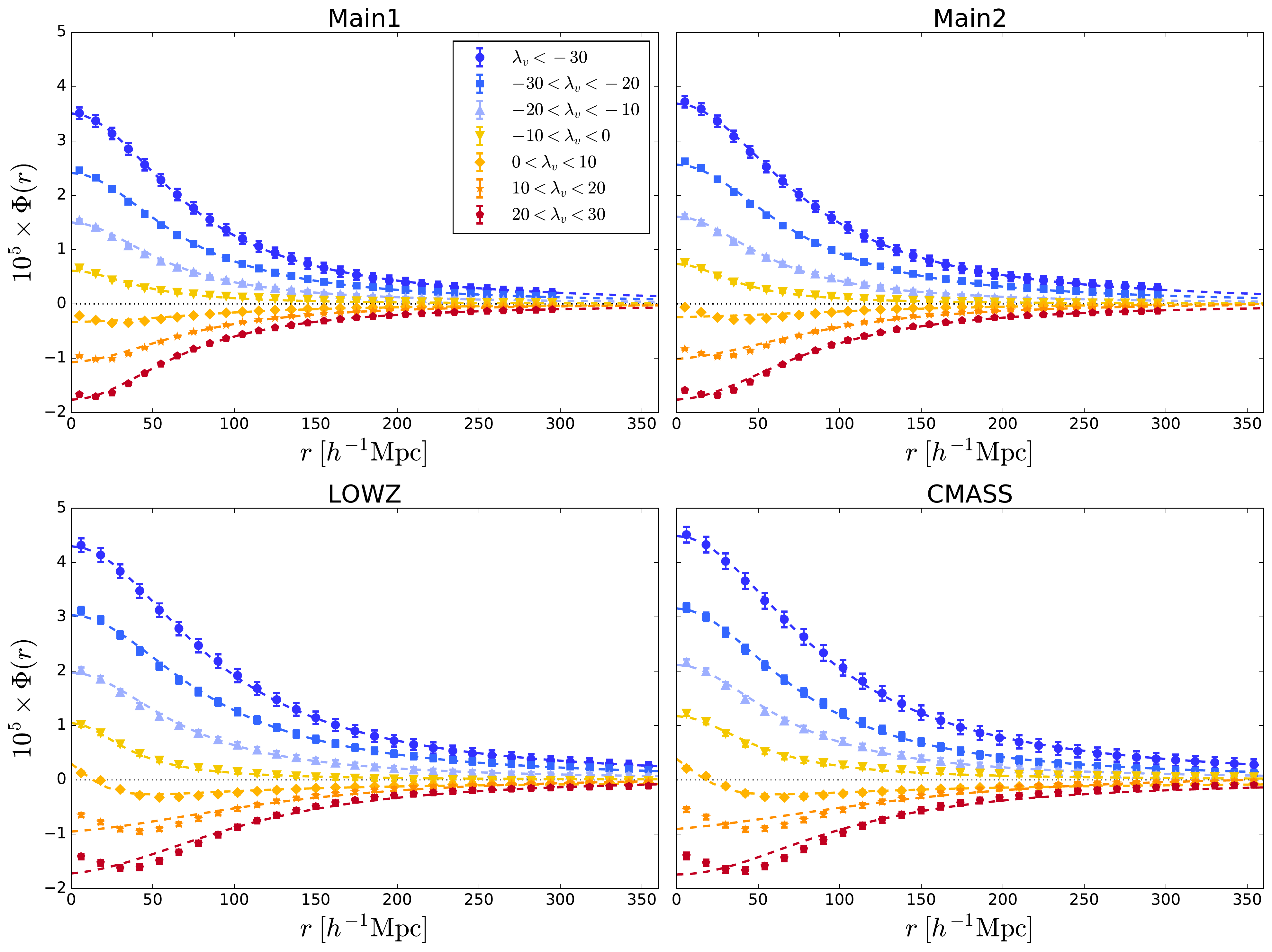}
\caption{Average gravitational potential profiles $\Phi(r)$ about the locations of void centres in each of the mock galaxy samples, for void stacks with different values of $\lambda_v$. Error bars show the $1\sigma$ errors in the mean. The dashed lines are fits to the data from equations \ref{eq:Phi_fit1} and \ref{eq:Phi_fit2}, for voids with $\lambda_v<0$ and $\lambda_v>0$ respectively. Voids with $\lambda_v>30$ have been omitted for clarity.} 
\label{fig:Phi_profiles}
\end{center}
\end{figure*}
%==================Fig.:=======================%

An ability to predict the expected gravitational potential around void locations would be an important tool for cosmological applications. To do this, we stacked the voids in our mock galaxy samples in bins of $\lambda_v$, and measured the average potential profile $\Phi(r)$ in concentric spherical shells about the void centres $\mathbf{X}_v$ in each bin. The results are shown in Figure~\ref{fig:Phi_profiles}. 

In each bin with $\lambda_v<0$, we found that the profiles were extremely well described by the two-parameter fitting function
\beq
\label{eq:Phi_fit1}
\overline\Phi(r,\lambda_v) = \frac{\Phi_{0v}(\lambda_v)}{1+\left(r/r_{0v}(\lambda_v)\right)^2}\;.
\eeq
The best-fit forms of this function are shown alongside the data in Figure~\ref{fig:Phi_profiles}. The fitted parameters $\Phi_{0v}$ and $r_{0v}$ are themselves linear functions of $\lambda_v$. Details of the fits are provided in Appendix \ref{appendix:fits}.

For stacks with $\lambda_v>0$, a slightly different 3-parameter functional form,
\beq
\label{eq:Phi_fit2}
\overline\Phi(r,\lambda_v) = \Phi_{0v}(\lambda_v)\frac{1-r/r_1(\lambda_v)}{1+\left(r/r_{0v}(\lambda_v)\right)^2}\;,
\eeq
was found to provide an adequate fit to the data, as also shown in Figure~\ref{fig:Phi_profiles}. Note that for these voids, $\overline\Phi(r)<0$ for almost the entire range of scales, with the possible exception of the void centre. 

An important feature apparent from Figure \ref{fig:Phi_profiles} is that the potential fluctuations associated with voids extend over scales of $200$-$300\;h^{-1}$Mpc, much larger than the physical sizes of the voids determined by the void-finding algorithm and shown in Figure \ref{fig:voidprops}. The reason for this can be understood from the form of equation \ref{eq:PoissonF}, which gives $\Phi(k) \propto k^{-2}\delta(k)$, enhancing the effect of long-wavelength modes. These long-wavelength modes are crucial to the understanding of the gravitational potential around voids. A practical consequence is that studies of voids from $N$-body simulations with too small a box length $L$, or from simulations in which long-wavelength modes are set to zero by hand \citep[e.g.,][]{Cai:2014,Cai:2016b}, fail to capture the full effect of these objects.

\subsection{Density profiles around void locations}
\label{subsec:density}

%==================Fig.: =======================%
\begin{figure*}
\begin{center}
\includegraphics[width=150mm]{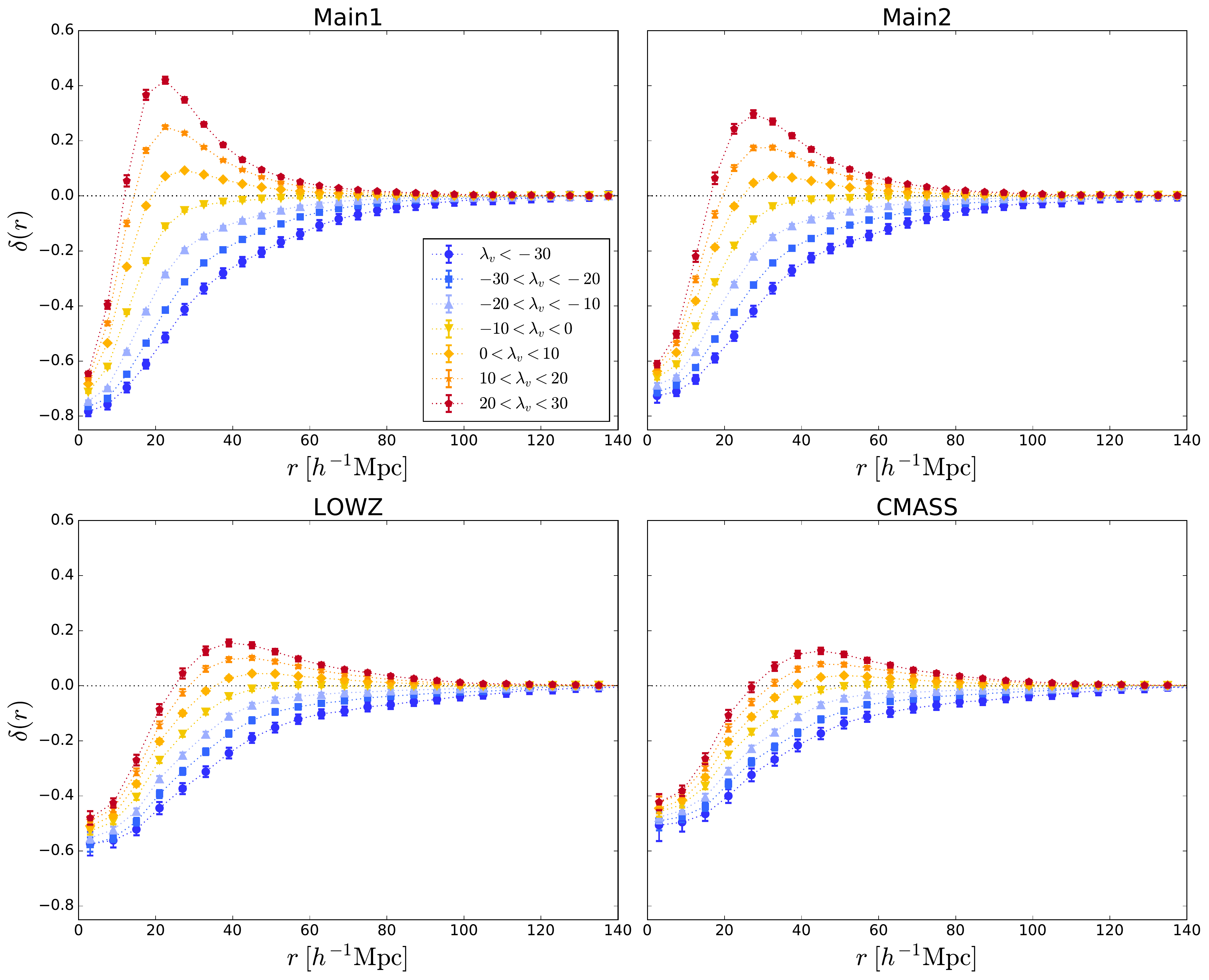}
\caption{Average density profiles $\delta(r)$ about the locations of void centres in each of the mock galaxy samples. Void stacks were created based on the $\lambda_v$ values as shown and match the corresponding stacks in Figure \ref{fig:Phi_profiles}. The density was measured in concentric shells of width $6\;h^{-1}$Mpc about the centre using the full simulation output on each redshift slice as described in the text. Error bars have been multiplied by a factor of 5 for visualisation purposes. Dotted lines are linear interpolations between the data points.} 
\label{fig:DM_profiles}
\end{center}
\end{figure*}
%==================Fig.:=======================%

The gravitational potential at void locations depends on the large-scale density environment around the voids. It should therefore be closely connected with the void density profiles. Since $\overline\Phi_0$ scales linearly with $\lambda_v$ in equation \ref{eq:Phiscaling}, this suggests that the void variable $\lambda_v$ should also be a useful proxy for separating out voids with \emph{under}-compensated central mass deficits from those \emph{over}-compensated voids with high-density surrounding walls.  

Figure \ref{fig:DM_profiles} shows the average profiles of the DM density $\delta(r)$ measured in spherical shells about the void centres, for stacks of voids selected according to measured values of $\lambda_v$. Note that this is determined from the full resolution simulation output and therefore is not the tracer number density profile that has been considered in some previous works \citep{Hamaus:2014a,Nadathur:2015a}. Tracer number density profiles within voids are harder to measure (see \citealt{Nadathur:2015a} for a discussion of some systematic effects) and are not simply related to the true DM density.

As expected, Figure \ref{fig:DM_profiles} shows that in all galaxy samples voids with large positive $\lambda_v$ show overdense walls surrounding the central underdensity, whereas those with large negative $\lambda_v$ do not. There is a steady trend towards increasing compensation with increasing $\lambda_v$, which is naturally related to the universal relationship for void compensation as a function of $\overline\delta_g$ found by \citet{Nadathur:2015c}. 

As the Main1 and Main2 mock samples are both drawn from the same simulation snapshot, the voids in these two populations trace the same DM density distribution. Therefore a comparison of the top two panels in Figure \ref{fig:DM_profiles} isolates the resolution effects of the higher-density tracer sample on the void identification: this results in slightly lower densities at void centres and better resolution of the high-density walls surrounding high-$\lambda_v$ voids. 

On the other hand, comparison of these profiles with those obtained from the LOWZ and CMASS void populations compare different redshift slices and thus show the effect of time evolution within the simulation. Importantly, since the velocity field $\mathbf{v}\propto-\nabla\Phi=0$ at the locations of maxima of the potential, these locations are \emph{stationary points} which do not shift within the simulation box as the time evolution progresses. Therefore, to the extent that voids with large negative values of $\lambda_v$ are effective tracers of these maxima of $\Phi$ as shown in Section \ref{subsec:maxima}, the profiles for such voids in the different panels of Figure \ref{fig:DM_profiles} show the time evolution of the density perturbation \emph{at the same locations}, from the $z=0.52$ snapshot (CMASS) to $z=0.1$ (Main1 and Main2). 

Another interesting observation is that despite the large differences in void size $R_v$ determined from the tracer galaxy distribution in the four mock galaxy samples apparent in Figure \ref{fig:voidprops}, the underlying DM density fluctuations that the voids correspond to extend over very similar scales, $\sim60\;h^{-1}$Mpc and typically larger than $R_v$. This is especially true for voids with large negative values of $\lambda_v$. This is because \zobov reports the void size based purely on the separation of regions based on the watershed transform, without reference to the absolute value of the density field. Thus a faint intervening density ridge within a large underdense void -- which is more likely to be resolved in high tracer density samples -- can lead to premature truncation of the watershed algorithm. This reduces the value of $\overline\delta_g$ (moving it towards larger negative values) and also biases $R_v$ low. Inferring the large-scale density environment from either of these observables individually is therefore complicated. Nevertheless, the particular combination of $R_v$ and $\overline\delta_g$ encoded with $\lambda_v$ is insensitive to this and thus provides a better universal indicator.

\subsection{Extension to overdensities}
\label{subsec:overdensities}

%==================Fig.: =======================%
\begin{figure}
\begin{center}
\includegraphics[width=85mm]{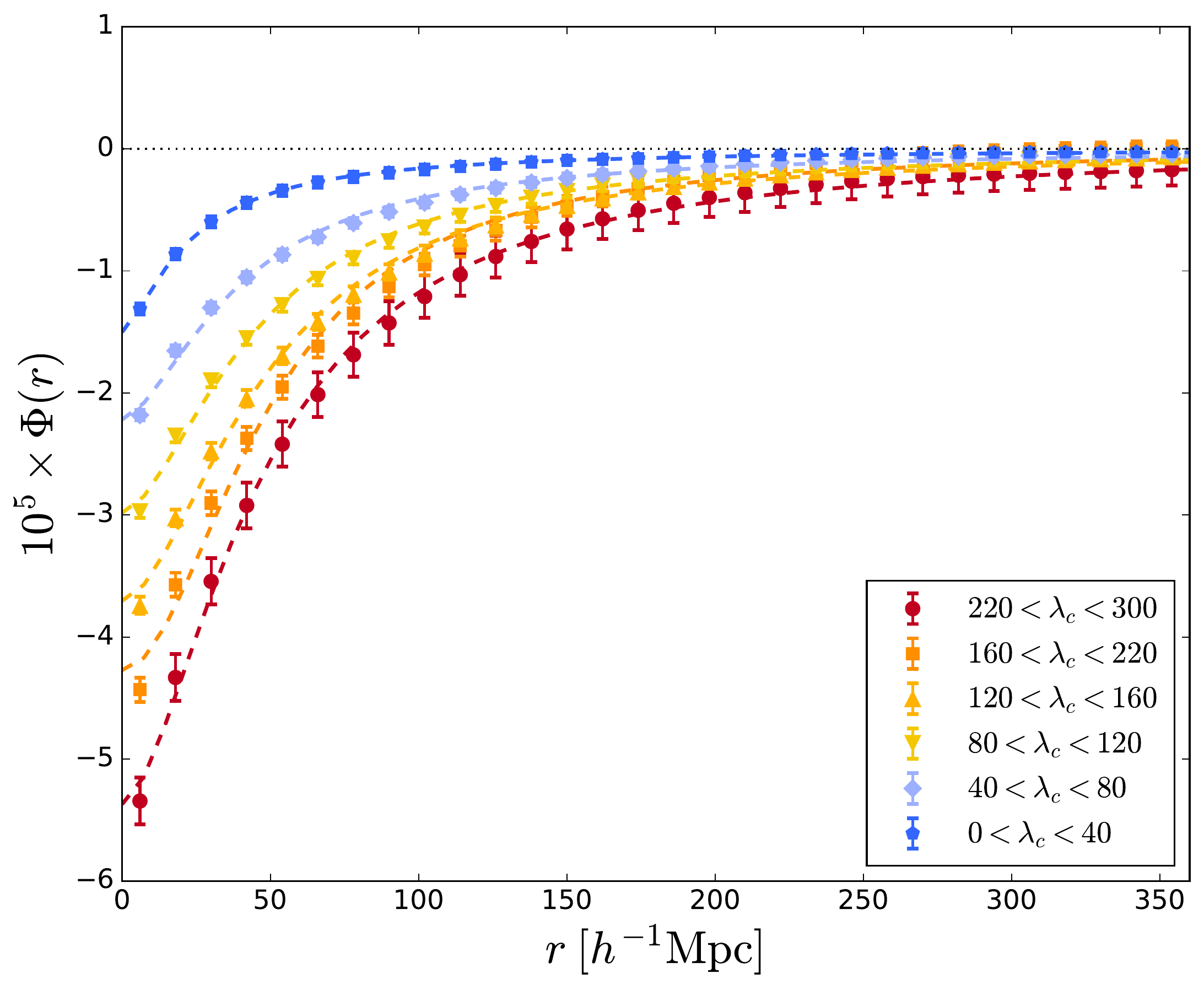}
\caption{Average gravitational potential profiles $\Phi(r)$ about the locations of superclusters in the mock CMASS galaxy sample. Superclusters are binned according to the value of $\lambda_c$ (see text). Dashed lines are fits to the data from equation \ref{eq:Phi_fit3}. Similar trends were found for $\Phi(r)$ profiles for superclusters in the other mock catalogues.} 
\label{fig:clust_Phi_profiles}
\end{center}
\end{figure}
%==================Fig.:=======================%

The \zobov algorithm we have used to identify voids can equally be inverted to instead find large overdense regions or `superclusters'. To do this we applied the same watershed algorithm to the inverse of galaxy density field reconstructed by the tessellation procedure, thus finding the locations of density maxima. As with voids, individual density maxima separated by watershed ridges were not merged to form larger structures. We defined the centre of each supercluster to be the position of the member galaxy with the smallest Voronoi cell (highest VTFE-reconstructed density). The superclusters thus obtained, analogously to the voids, do not necessarily correspond to collapsed or gravitationally bound individual structures, rather to DM density fluctuations extending over large scales. Such supercluster regions have been used in ISW measurements \citep{Granett:2008a,Hotchkiss:2015a,Nadathur:2016b,Kovacs:2016}.

Superclusters in the simulated galaxy populations are much more numerous and also typically smaller than the corresponding set of voids. However, they trace the locations of minima of the gravitational potential in a similar way to voids tracing maxima. Analogously to equations \ref{eq:Phiscaling} and \ref{eq:lambda_v}, we found that the gravitational potential at the supercluster centres was well predicted by the quantity
\beq
\label{eq:lambda_c}
\lambda_c \equiv \overline\delta_g\left(\frac{R_c}{h^{-1}\rmn{Mpc}}\right)^{1.6}\;,
\eeq
where $R_c$ is the effective spherical radius of the supercluster and $\overline\delta_g$ is defined as for voids. (Note the difference in the exponents in equations~\ref{eq:lambda_c} and \ref{eq:lambda_v}.) A linear scaling relationship between $\overline\Phi_0$ and $\lambda_c$ was found to hold for superclusters as well. The stacked potential profile around supercluster locations was found to be well described by a fitting formula similar to equation \ref{eq:Phi_fit1}, 
\beq
\label{eq:Phi_fit3}
\overline\Phi(r,\lambda_c) = \frac{\Phi_{0c}}{1+\left(r/r_{0c}\right)^\alpha}\;,
\eeq
with $\Phi_{0c}(\lambda_c)$, $r_{0c}(\lambda_c)$ and $\alpha(\lambda_c)$ all fit to the data from simulation. An example of the measured $\Phi(r,\lambda_c)$ profiles and the fits are shown in Figure \ref{fig:clust_Phi_profiles} for superclusters in the CMASS sample.

\section{Applications}
\label{sec:applications}

In the following we discuss some applications of the empirical results obtained above for the understanding of the nature of cosmic voids and their use in precise cosmological tests.

\subsection{ISW effect of superstructures}
\label{subsec:ISW}

One example of the use of these results is in attempts to measure the secondary anisotropies in the CMB introduced by the ISW effect of cosmic structures. In the linear approximation which is valid on the scales of interest \citep[e.g.,][]{Cai:2010hx,Nadathur:2014b}, the ISW temperature shift in the CMB temperature induced along a direction $\mathbf{\hat n}$ can be expressed in terms of fluctuations in the gravitational potential:
\beq
\label{eq:ISWint}
\frac{\Delta T_\rmn{ISW}}{\overline{T}}(\mathbf{\hat n}) = -2\int a(z)\left(1-f(z)\right)\Phi\left(\mathbf{\hat n},z\right)\,\rmn{d}z\;,
\eeq
where $f= \frac{\rmn{d}\ln D}{\rmn{d}\ln a}$ is the growth rate of structure and the integral is over redshift $z$. 

The results of this paper allow the development of a strategy to detect the ISW contributions from individual voids and superclusters using a stacking analysis. This strategy has the following elements:
\begin{enumerate}
\item The locations of (subsets of) voids and superclusters found in galaxy survey data allow identification of the locations of maxima and minima of $\Phi$, respectively, which cause negative and positive ISW temperature shifts.
\item The observable quantities $\lambda_v$ and $\lambda_c$ can be used to characterise these potential fluctuations, and thus to order structures by the magnitude of the temperature effects they produce: the linear scaling found in Sections \ref{subsec:scaling} and \ref{subsec:overdensities} means that the expected temperature $\overline{\Delta T_{v,c}}$ also scales linearly with the variables $\lambda_{v,c}$ for both voids and superclusters.
\item The fitting formulae \ref{eq:Phi_fit1}, \ref{eq:Phi_fit2} and \ref{eq:Phi_fit3} allow precise predictions for the sky profile $\overline{\Delta T(\theta)}$ around these points of extrema, and thus can be used to construct optimal matched filters with which to separate the small late-time ISW contribution from the background of primordial CMB fluctuations.
\end{enumerate}
A detailed discussion of this method has recently been provided by \citet{Nadathur:2016b}, who showed that it has a sensitivity comparable to the traditional cross-correlation approach. Using data from \emph{Planck} \citep{Planck:2015Overview} and a catalogue of voids and superclusters drawn from the SDSS Data Release 12 CMASS galaxy sample \citep{Nadathur:2016a}, they were able to obtain a high-significance detection of the ISW effect in excellent agreement with predictions derived from the simulations considered in this work.

\subsection{Void lensing}
\label{subsec:lensing}

%==================Fig.: =======================%
\begin{figure}
\begin{center}
\includegraphics[width=80mm]{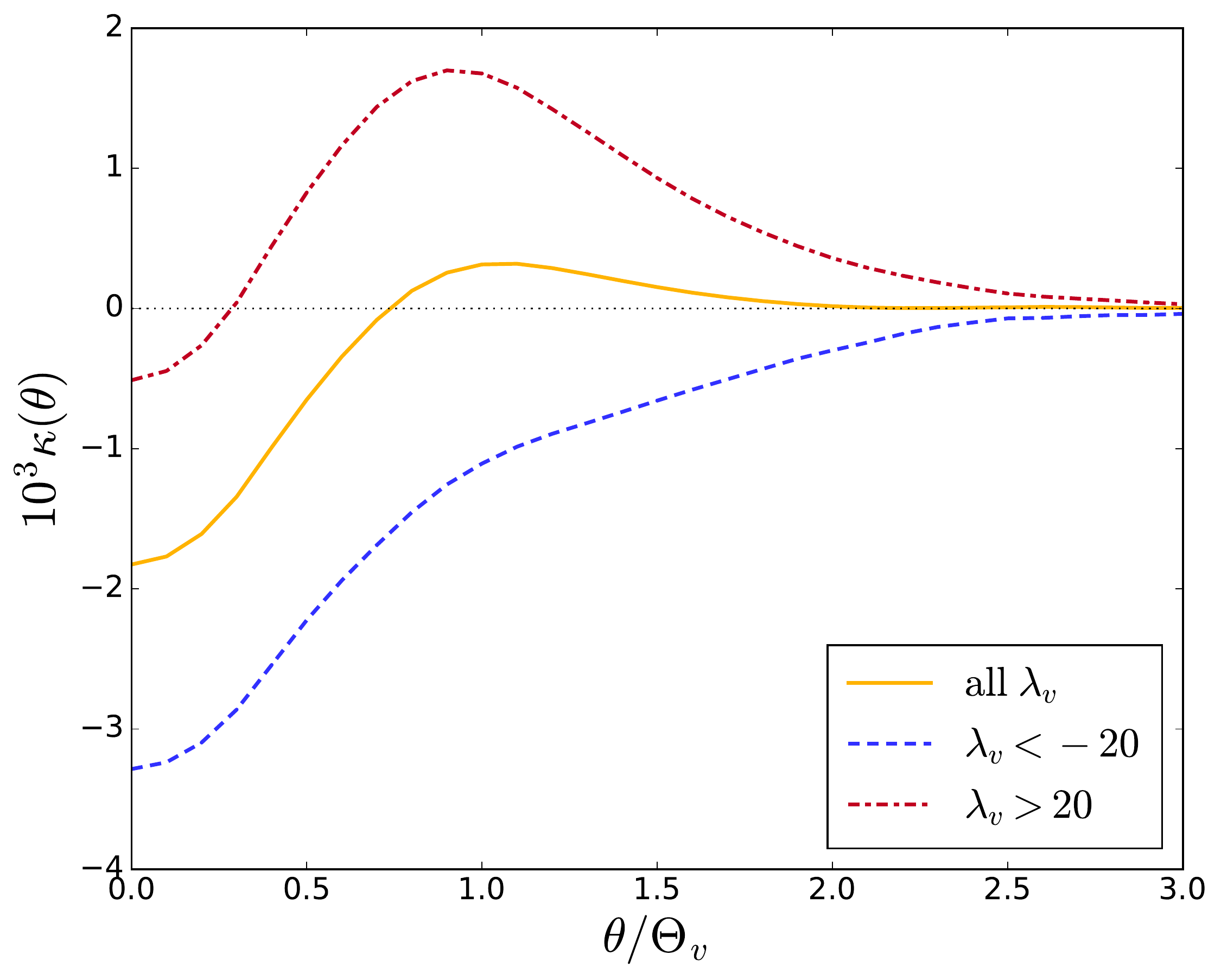}
\caption{The yellow (solid) line shows the average CMB lensing convergence profile $\kappa(\theta)$ for all voids with $40\;h^{-1}\rmn{Mpc}<R_v<60\;h^{-1}$Mpc in the CMASS mock catalogue. Angular units are scaled in terms of the average angular size of these voids, assuming they are centred at redshift $z=0.52$. The blue (dashed) and red (dot-dashed) lines show $\kappa(\theta)$ for two subsets of this sample, with $\lambda_v<-20$ and $\lambda_v>20$ respectively but with the same average void size. For typical numbers of voids in survey data, statistical uncertainties in these predictions will be much smaller than observational errors, so are omitted here.} 
\label{fig:kappa}
\end{center}
\end{figure}
%==================Fig.:=======================%

Measuring the gravitational lensing effects of voids on background source galaxies has been the subject of several recent studies \citep{Krause:2013,Melchior:2014,Clampitt:2015,Sanchez:2016}. Recently, \citet{Cai:2016b} have also made the first measurement of the effects of voids on the CMB lensing convergence $\kappa$. Precise measurements of the lensing signal of voids may allow a detailed mapping of the dark matter distribution within them, which it is hoped may be useful as a test of modified gravity models \citep[e.g.][]{Clampitt:2013,Cai:2015,Barreira:2015}.

Void lensing studies to date have characterised the expected lensing effect of voids purely on the basis of the void size $R_v$. However, as the lensing potential is sourced by the gravitational potential $\Phi$, and the void parameter $\lambda_v$ is a useful proxy for $\Phi$, it follows that $\lambda_v$ should also provide a useful discriminant between populations of voids that have the same size $R_v$ yet produce very different lensing effects. To illustrate this, we used the stacked average DM density profile $\delta(r)$ for voids, determined as in Section \ref{subsec:density}, to calculate the CMB lensing convergence signal,
\beq
\label{eq:kappa}
\kappa(\theta)=\frac{3\Omega_\rmn{m}H_0^2}{2c^2}\int \frac{\chi\left(\chi_s-\chi\right)}{\chi_s}\frac{\delta(\theta,\chi)}{a}\rmn{d}\chi,
\eeq  
where $\chi$ is the comoving radial coordinate and $\chi_s$ is the comoving distance to the last scattering surface. Figure \ref{fig:kappa} shows the resultant average $\kappa(\theta)$ signal for all voids in the CMASS mock void catalogue in the size range $40\;h^{-1}\rmn{Mpc}<R_v<60\;h^{-1}$Mpc as the yellow solid line. Also shown are the $\kappa(\theta)$ profiles for two additional subsets of voids, which both satisfy exactly the same size cuts, but have $\lambda_v<-20$ (blue dashed) and $\lambda_v>20$ (red dot-dashed) respectively.

It is clear that voids of very similar size $R_v$ but different mean galaxy density $\overline\delta_g$ and thus $\lambda_v$ can produce very different lensing convergence signals. Equally, as the apparent size $R_v$ is only loosely related to the true extent of the void DM underdensity (Section \ref{subsec:density}), voids with very different $R_v$ could contribute similar convergence profiles $\kappa(\theta)$. In addition, Figure \ref{fig:kappa} shows that averaging together the contributions from voids with different values of $\lambda_v$ will in general produce an average convergence that is closer to zero and thus potentially harder to measure. This suggests that the sensitivity of detection of void lensing effects could be significantly improved by consideration of sub-populations defined by the variable $\lambda_v$.

Although we have only discussed the convergence $\kappa$ in the example above, the same argument can equally be applied to the contribution of voids to the lensing shear $\gamma$. We leave further exploration of these effects and applications to data to future work.

\subsection{Voids in redshift space}
\label{subsec:RSD}

Voids in galaxy surveys are observed in redshift space. Under the assumption of an isotropic Universe the stacked galaxy distribution around void centres should average to spherical symmetry in real space, but will in general appear distorted due to the Alcock-Paczynski (AP) effect \citep{Alcock:1979}. This has been proposed as a potentially powerful test of cosmology \citep{Lavaux:2012}, which has recently been applied to galaxy survey data \citep{Sutter:2012tf,Sutter:2014d,Mao:2016b,Hamaus:2016}.

The use of voids for the AP test is complicated by redshift-space distortions due to peculiar velocities. Naively, one would expect velocity outflows around voids, leading to a stretching of their shapes along the line of sight when seen in redshift space. However, several authors \citep{Lavaux:2012,Sutter:2014d,Mao:2016b} have found the opposite: seen in redshift space, voids identified using watershed void-finders such as \zobov instead appear to be \emph{squashed} along the line of sight. This phenomenon has been noted both in simulations and for voids in real galaxy data. \citet{Mao:2016b} describe it as a failure of linear theory and show that it degrades the sensitivity of the AP test. However, \citet{Cai:2016a} argue that a squashing effect can be consistent with linear theory.

Our results provide another perspective: voids reside in a variety of different large-scale environments, so not all voids are associated with velocity outflows. As noted in Section \ref{subsec:inferringPhi}, a significant fraction of voids that are identified by the watershed algorithm correspond to local density minima within regions that are overcompensated on large scales and thus form potential wells rather than maxima. In linear theory, such regions will not correspond to velocity outflows (at least on scales of observational interest). Another way to illustrate the same problem is to note that a dynamical method of classification of the cosmic web based on eigenvalues of the tidal tensor $\mathbf{T}_{\alpha\beta}=\partial_\alpha\partial_\beta\Phi$ \citep{Hahn:2007} shows that only a small fraction of the volume of the Universe should lie in regions that are simultaneously expanding along all three directions, whereas watershed void-finders such as \zobov are by nature space-filling. Only a fraction of \zobov voids can correspond to local maxima of $\Phi$ and thus to truly expanding regions. 

In addition, on the basis of our results we can make a few qualitative predictions. Firstly, the minority of voids with large negative values of $\lambda_v$ should correspond to strong velocity outflows. Secondly, the magnitude of the velocity outflow and thus the details of the induced redshift-space distortion should vary with the values of $\lambda_v$, as should the length scale over which the effect is observable. Thirdly, as discussed in Appendix \ref{appendix:centres}, the void centre $\mathbf{X}_v$ used in this work traces maxima of $\Phi$ better than the void barycentre $\mathbf{X}_b$ used in many previous analyses, and therefore a shift from  
use of $\mathbf{X}_b$ to $\mathbf{X}_v$ in void catalogues will enhance the velocity outflow seen. Detailed exploration of these topics and application to voids observed in real galaxy data is left to future work.

\subsection{Constructing a void number function}
\label{subsec:number_func}

One goal of void studies is to obtain a prediction for the void number function, analogous to the halo mass function. Following \citet{Sheth:2003py}, theoretical models for the void number function based on the excursion set formalism attempt to predict the abundance of voids as a function of their effective size, $R_v$, alone. However, the \citet{Sheth:2003py} model fails by orders of magnitude to predict the void size distribution observed in simulation or in galaxy data. Various modifications of the model can be tuned to match the large-$R_v$ cutoff, but still fail to describe the full observed distribution, and also fail to match other predictions for void properties \citep[for a fuller discussion, see][]{Nadathur:2015b}.

An important reason for this discrepancy is the difficulty in modelling the action of watershed void-finding algorithms on the late-time, non-linear density field, especially as it is traced by the discrete distribution of biased galaxies. The model makes strong assumptions about spherical evolution of voids until the point of shell-crossing in order to obtain a threshold value in the linearised density field, $\delta_\rmn{v}=-2.71$, which is used to define voids and to set the excursion set barrier; these assumptions are however not satisfied by watershed voids. The dependence of apparent void size on the tracer galaxy properties discussed in Section \ref{subsec:density} adds a further layer of complexity.

Our results suggest a new approach to the problem. Instead of modelling the non-linear dark matter density field, an alternative starting point would be to predict the number density of peaks in the smoothed gravitational potential field. This is mathematically a far simpler task \citep{BBKS}, as the $\Phi$ field remains in the linear regime and very close to Gaussian even at late times. Predictions for peaks of $\Phi$ could then be related to the mean void values of $\lambda_v$ according to the empirical fits provided in this work.  

Such an approach would involve a fundamental shift in the concept of the void number function: according to this view, it would describe the abundance of voids according to the \emph{combination} of their size and average density, $\lambda_v$, rather than as a function of size alone. A fuller exploration of this approach is left to future work.

\section{Discussion}
\label{sec:discussion}

Distinguishing the physical environment around voids is of key importance both to developing them as probes of fundamental physics, as well as to understanding their formation and evolution. Our aim in this paper has been to examine how cosmic voids, which are in practice identified as local minima of the galaxy density field, relate to the gravitational potential $\Phi$. This is a direct continuation of the results of \citet{Nadathur:2015b} and \citet{Nadathur:2015c}, where we examined the relationship between galaxy voids and the underlying matter density field $\delta$. 

The results we have presented allow us to answer the questions posed in Section \ref{sec:introduction} as follows. A significant fraction of galaxy voids -- in some cases, even a majority -- do not lie in expanding regions and do not correspond to $\Phi>0$. This can be easily understood, as voids simply correspond to local minima of the density field but $\Phi$ depends on the large-scale distribution of mass around such minima. 

Nevertheless, a substantial subset of all voids \emph{does} trace the positions of maxima of $\Phi$. Such voids can be identified on the basis of the void parameter $\lambda_v$, which combines information on the average underdensity of the void $\overline\delta_g$ and its apparent spatial extent $R_v$, and can be easily measured in practice. The value of $\lambda_v$ is a good predictor of the value of $\Phi$ at the void location, with a very simple universal linear scaling between the two quantities for voids traced by any galaxy type and at any redshift.  In fact, the spherically averaged profiles $\Phi(r)$ about void locations are also very well determined by the value of $\lambda_v$. 

All of our results have been purely empirically obtained from simulations. Developing a theoretical model of such voids from first principles requires a proper modelling of the action of the \zobov watershed algorithm on reconstructed density fields, a task that appears too complex for an analytic solution. In this situation, our empirical findings, and particularly the linear scaling between $\lambda_v$ and $\Phi$, might provide a better starting point for such a project. 

At a more immediate pragmatic level, the results of this paper have important consequences for the use of voids in cosmological tests, particularly for measurement of the ISW and lensing signals of voids, and for the estimation of velocity flows around void locations. Indeed the practical application to ISW measurements has already been demonstrated in \citet{Nadathur:2016b}, where we used the fits to $\Phi(r)$ obtained here to devise to new sensitive matched-filtering technique to extract the tiny ISW signal of voids and superclusters from primordial CMB noise. We anticipate that our results will be similarly useful in other areas.

It is however worth noting that the empirical fits obtained here cannot be assumed to apply with the same precision for voids defined according to alternative void-finding algorithms, or for other treatments of void merging (see Appendix~\ref{appendix:centres}). The use of photometric redshift data, such as from the Dark Energy Survey (DES)\footnote{\url{http://www.darkenergysurvey.org/}}, the Large Synaptic Survey Telescope (LSST)\footnote{\url{http://www.lsst.org/}} or from Euclid\footnote{\url{http://www.euclid-ec.org/}}, will also affect the efficiency of detection of voids and superclusters, as well as potentially biasing the reconstruction of their sizes and densities, due to the smearing effect of large photometric redshift uncertainties. These effects mean that the results obtained here might require adjustment for use with photometric data. We leave a detailed study of these effects to future work.

\section{Acknowledgements}

We thank Yanchuan Cai and John Peacock for helpful comments. S.N. acknowledges an Individual Fellowship of the Marie Sk\l odowska-Curie Actions under the H2020 Framework of the European Commission, project {\small COSMOVOID}. RC is supported by the UK Science and Technologies Facilities Council grant ST/N000668/1. The BigMultiDark simulations were performed on the SuperMUC supercomputer at the LeibnizRechenzentrum in Munich, using resources awarded to PRACE project number 2012060963. We acknowledge use of the {\small EREBOS}, {\small THEIA} and {\small GERAS} clusters at the Leibniz Institut f\"ur Astrophysik (AIP).

\bibliographystyle{mn2e}
\bibliography{refs.bib}

\appendix
\section{Effect of void centre definition and void merging}
\label{appendix:centres}

%==================Fig.: =======================%
\begin{figure*}
\begin{center}
\includegraphics[width=150mm]{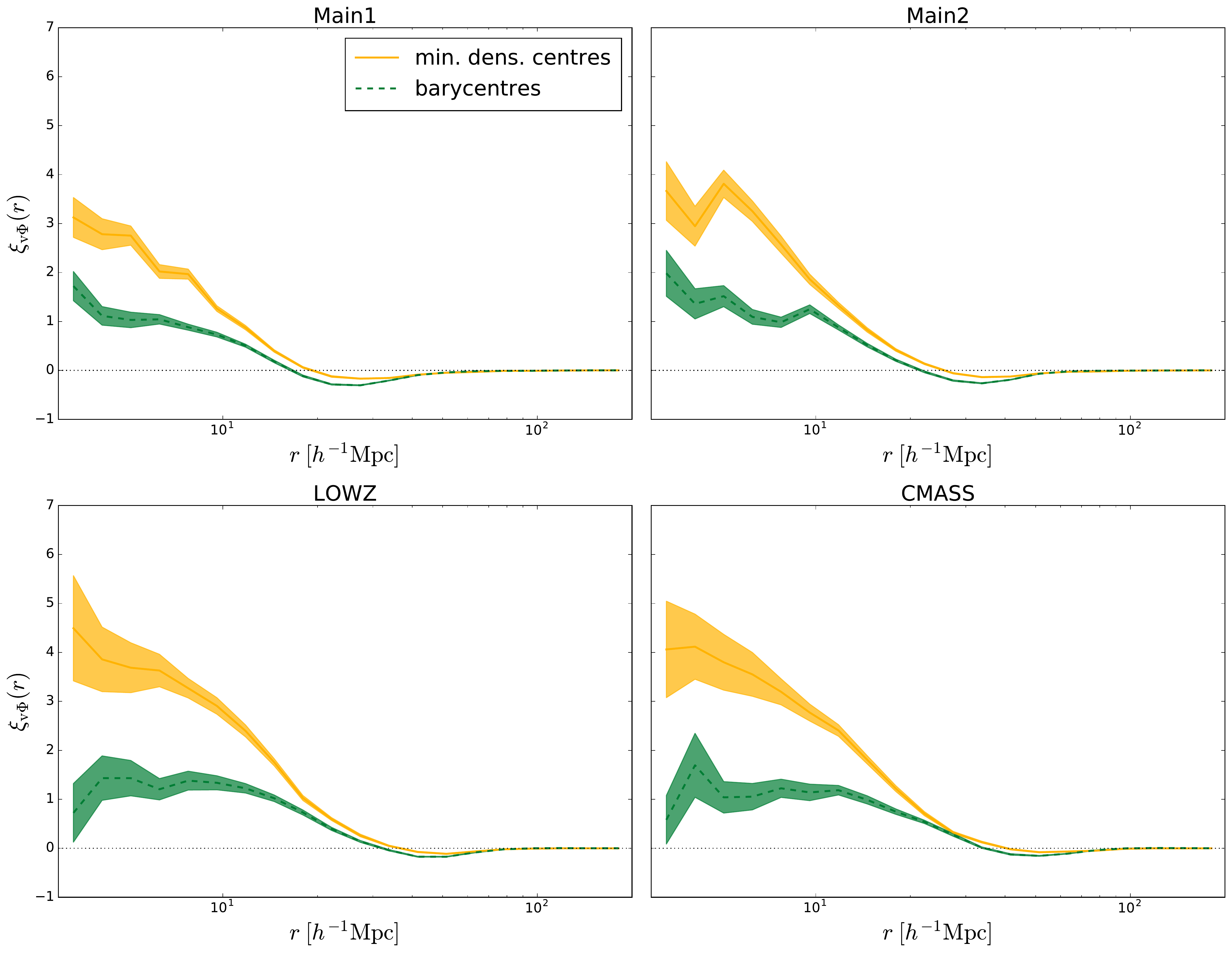}
\caption{Effect of void centre definition on the correlation between voids and maxima of the gravitational potential. The yellow solid line shows the cross-correlation $\xi_{v\Phi}(r)$ between void locations $\mathbf{X}_v$ and maxima of $\Phi$ for all voids as in Figure~\ref{fig:CCF_all}. The green dashed line shows the cross-correlation for the same voids, but when the void location is defined by the weighted average of the position of galaxies within the void, $\mathbf{X}_b$, instead. Shaded bands indicate the $1\sigma$ confidence regions estimated from jack-knife errors in both cases. The centre locations $\mathbf{X}_v$ used in this work provide a much better tracer of maxima of $\Phi$ for voids in all mock samples.} 
\label{fig:CCF_cvsb}
\end{center}
\end{figure*}
%==================Fig.:=======================%

The location of the `centre' of an irregular-shape void is ambiguous. We have so far defined the void centre location to be the location $\mathbf{X}_v$ corresponding to the minimum of the galaxy density field. An alternative centre position may be defined as the weighted centroid of the positions of galaxies within the void. This is often referred to as the void `barycentre', $\mathbf{X}_b$, and has been commonly used in a number of void studies.

In order to compare the relative merits of the two centre definitions $\mathbf{X}_v$ and $\mathbf{X}_b$ as tracers of the gravitational potential, we recomputed the cross-correlation function $\xi_{v\Phi}$ as in Section~\ref{subsec:maxima} for the barycentre definition. A comparison of the results obtained for the two definitions is shown in Figure \ref{fig:CCF_cvsb}. It is clear that in all cases the void minimum density centre $\mathbf{X}_v$ is significantly better correlated with maxima of $\Phi$ and thus provides a better tracer of the gravitational potential. This is consistent with earlier results \citep{Nadathur:2015b,Nadathur:2015c} showing that $\mathbf{X}_v$ is also a significantly better tracer of the true minimum of the dark matter density field traced by galaxy voids.

As discussed in Section \ref{subsec:void-finding}, we do not merge neighbouring voids together to form larger voids. Such merging of voids has commonly been used in other studies in the literature, but the choice of criteria to govern when merging occurs is necessarily always subjective \citep[see][]{Nadathur:2015c}, and different authors have used widely differing prescriptions. It is not possible to test all such choices, nor do we attempt a comprehensive comparison. Instead we tested a few representative examples of choices for void merging, as described by \citet{Nadathur:2015b}. In all cases, the cross-correlation between void centres and maxima of $\Phi$, $\xi_{v\Phi}$, was reduced relative to the values obtained without merging. The decrease in the correlation was found to be smaller when the criteria for merging were tightened, resulting in fewer merged voids. We conclude that the merging of neighbouring voids degrades their use as tracers of the gravitational potential. Combined with the reduction in statistical power and the disadvantages of merging already noted by \citet{Nadathur:2015c}, this leads us to recommend that void merging \emph{not} be used in void studies.

Irrespective of the choice of the merging criteria investigated, the correlation $\xi_{v\Phi}$ with peaks of the gravitational potential was significantly higher for void minimum density centres $\mathbf{X}_v$ than barycentres $\mathbf{X}_b$, as in the case shown in Figure ~\ref{fig:CCF_cvsb}. 

\section{Numerical fits to data}
\label{appendix:fits}

Table \ref{table:Phi0fit} shows the numerical fits to equation \ref{eq:Phiscaling} for the gravitational potential at the void centre, $\overline\Phi_0(\lambda_v)$ for voids in all four mock galaxy samples. A simple linear scaling provides an excellent fit for $\overline\Phi_0$ in all cases. Voids with all values of $\lambda_v$ were included in obtaining the fits. 

Also included in Table \ref{table:Phi0fit} are the mean standard deviations $\sigma_{\Phi_0}$ for the distributions of values of $\Phi_0$ for individual voids. This quantity does not show any significant dependence on $\lambda_v$. The values of $\sigma_{\Phi_0}$ indicate a large void-to-void scatter. Therefore precise predictions for $\Phi_0$ can only be made for the average over several void locations and not for individual voids.

The slopes $a$ in Table \ref{table:Phi0fit} are somewhat smaller for the two low-bias Main galaxy mocks than for the highly biased LOWZ and CMASS mocks. This difference was much reduced when only voids with $\lambda_v<0$ were included in the fit, indicating that it arises due to the greater number of overcompensated voids in high-density environments that are resolved in the Main galaxy samples. 

\begin{table}
\centering
\caption{Numerical fits for $\overline\Phi_0(\lambda_v)$ and the mean standard deviation of the distribution of residuals}
\begin{tabular}{@{}cccc}
\hline
Sample &  \multicolumn{2}{c}{$\overline\Phi_0 = -a\lambda_v + c$} & $\sigma_{\Phi_0}$ \\
 & \multicolumn{2}{c}{-----------------} & \\
 & $a\times10^{5}$ & $c\times10^{5}$ & $\times10^{5}$ \\
\hline
 Main1 & $0.083\pm0.001$ & $0.316\pm0.007$ & 3.20\\
Main2 & $0.082\pm0.001$ & $0.443\pm0.009$ & 3.14\\
LOWZ & $0.090\pm0.001$ & $0.731\pm0.016$ & 3.30\\
CMASS & $0.091\pm0.001$ & $0.853\pm0.019$ & 3.42\\
\hline
\end{tabular}
\label{table:Phi0fit}
\end{table}

In addition, Table \ref{table:Phi_r_fit} shows details of the fits to the stacked profiles $\Phi(r)$ about void centres. For simplicity, we report profile fits only for $\Phi_{0v}(\lambda_v)$ and $r_{0v}(\lambda_v)$ in equation \ref{eq:Phi_fit1}, applicable to voids with $\lambda_v<0$, since such voids are anticipated to be more useful in detection of cosmological signals such as the ISW effect \citep{Nadathur:2016b}. Both these variables are also approximately linear functions of $\lambda_v$.

 \begin{table}
\centering
\caption{Fitted values of $\Phi_{0v}$ and $r_{0v}$ for voids with $\lambda_v<0$}
\begin{tabular}{@{}ccccc}
\hline
Sample &  \multicolumn{2}{c}{$\Phi_{0v} = -a\lambda_v + c$} & \multicolumn{2}{c}{$r_{0v} = -m\lambda_v + b$}\\
 & \multicolumn{2}{c}{-----------------} & \multicolumn{2}{c}{-----------------} \\
 & $a\times10^{5}$ & $c\times10^{5}$ & $m$ & $b$\\
\hline
 Main1 & $0.095\pm0.001$ & $0.22\pm0.01$ & $1.35\pm0.04$ & $37.7\pm0.8$\\
Main2 & $0.095\pm0.001$ & $0.31\pm0.01$ & $1.65\pm0.05$ & $33.0\pm1.0$\\
LOWZ & $0.106\pm0.002$ & $0.53\pm0.03$ & $1.94\pm0.07$ & $33.1\pm1.3$\\
CMASS & $0.103\pm0.002$ & $0.70\pm0.03$ & $1.29\pm0.08$ & $46.8\pm1.6$\\
\hline
\end{tabular}
\label{table:Phi_r_fit}
\end{table}

\label{lastpage}
\end{document}